\documentclass{article}

\usepackage[utf8]{inputenc}
\usepackage[a4paper, total={6in, 8in}]{geometry}

\usepackage{amsmath,amssymb,bm}
\usepackage{graphicx}
\usepackage{float}
\usepackage{placeins}
\usepackage{listings}
\usepackage{xcolor}
\usepackage{url}
\usepackage[affil-it]{authblk}

\usepackage[nottoc,numbib]{tocbibind}

\usepackage{hyperref}
\hypersetup{colorlinks,linkcolor={blue},citecolor={blue},urlcolor={blue}}

\usepackage[]{lineno}

\usepackage[style=phys]{biblatex}
\title{Unperturbed-orbit integration and the 3D kinetic dispersion relation of the electron cyclotron drift instability}

\author[1]{Yinjian Zhao\thanks{Corresponding author: Yinjian Zhao,
\href{mailto:zhaoyinjian@hit.edu.cn}{zhaoyinjian@hit.edu.cn}}}

\affil[1]{School of Energy Science and Engineering,
Harbin Institute of Technology,
Harbin 150001, People's Republic of China}

\date{\today}

\begin{document}
\maketitle

\begin{abstract}
High-frequency instabilities in crossed-field ($\bm E\times\bm B$) plasmas are widely implicated in
anomalous cross-field electron transport in Hall thrusters and related devices. Building on the
fully kinetic 3D electrostatic dispersion relations reported by Ducrocq \emph{et al.} and later by
Lafleur \emph{et al.}, we provide a concise, self-contained derivation of the key missing step: the
magnetized-electron density perturbation $n_{e1}$ obtained from the linearized Vlasov equation via a
retarded integration along unperturbed orbits, including finite-Larmor-radius effects and cyclotron
harmonics. We collect the required mathematical identities in appendices and clarify the mapping
between the Ducrocq Poisson-form and the Lafleur dielectric-form representations, including ion
closures (cold-fluid versus kinetic Landau response). We conclude with a brief discussion of the
assumptions and possible extensions toward more realistic configurations.
\end{abstract}

\section{Introduction}\label{sec:intro}

Crossed-field ($\bm E\times\bm B$) plasmas are ubiquitous in both fundamental plasma physics and
high-impact applications. Prominent examples include Hall effect thrusters for electric propulsion
\cite{Lafleur},
Penning-type discharges and magnetrons
\cite{penning}, as well as a broad class of partially magnetized plasma
devices used in plasma processing and plasma sources. In these systems, an imposed electric field
perpendicular to a magnetic field drives an azimuthal electron drift, while ions are typically
weakly magnetized or effectively unmagnetized. The resulting separation of magnetization between
species yields rich wave-particle dynamics and enables a variety of microinstabilities.

A central challenge in $\bm E\times\bm B$ plasmas is the long-standing problem of
\emph{anomalous} cross-field electron transport. In many practical regimes, measured electron
mobility across magnetic field lines exceeds classical collisional predictions by orders of
magnitude. A widely accepted physical picture is that high-frequency electrostatic instabilities
generate turbulence and correlated fluctuations that enhance electron momentum and energy exchange,
thereby producing an effective ``anomalous conductivity''.
Among these instabilities, the electron cyclotron drift instability (ECDI) and its long-wavelength
extensions (often discussed in connection with modified two-stream physics) play a particularly
important role in Hall thruster-like parameter ranges.

A quantitative understanding of these instabilities requires a kinetic description of magnetized
electrons, including finite Larmor radius (FLR) effects and cyclotron harmonics, coupled to an ion
response that may range from cold-fluid to kinetic (Landau) behavior. Ducrocq \emph{et al.}
\cite{Ducrocq} were the first to derive a fully three-dimensional (3D) electrostatic kinetic
dispersion relation for this class of crossed-field configurations. Later, Lafleur \emph{et al.}
\cite{Lafleur} presented an equivalent 3D formulation in a dielectric form and emphasized its
utility for modeling anomalous transport in Hall thrusters, including the option of retaining a
kinetic (unmagnetized) ion response. Despite the importance of these results, both derivations are
notoriously lengthy: only the key intermediate steps are reported in the original papers, which
makes the full logic difficult to reproduce for readers encountering the theory for the first time.

The primary purpose of the present work is therefore pedagogical. We provide a comprehensive and
self-contained derivation of the magnetized-electron density perturbation $n_{e1}$ starting from
the electron Vlasov equation and proceeding through the retarded (causal) integration along
unperturbed particle orbits. This particular step---from the linearized Vlasov equation to a
closed-form expression for $n_{e1}$---is the technical core of the full 3D dispersion relation and
is also the point where several standard but nontrivial identities enter (gyrophase dynamics,
Jacobi--Anger expansions, the Landau prescription, and plasma dispersion-function manipulations).
For ease of study, the appendices collect and derive the main mathematical ingredients required to
follow the calculation in detail.

After completing the derivation of the electron response in our notation, we clarify how the
Ducrocq ``Poisson-form'' dispersion relation and the Lafleur ``dielectric-form'' dispersion
relation are algebraically equivalent, and we explicitly identify the notational mappings (thermal
speed conventions, harmonic-sum layouts, and ion closures).
Finally, we discuss the main assumptions underlying the present derivation (uniform crossed fields,
electrostatic perturbations, linear response, and simplified equilibrium closures) and outline
possible extensions, including more general equilibria, additional physics (e.g., collisions or
electromagnetic effects), and directions for future theoretical development.

\section{Derivation of Electron Density Perturbation
}\label{sec:derivation}

\subsection{Model setup and linearization}
\label{subsec:setup_linear}

We follow the same model setup as described in
Ducrocq et al.~\cite{Ducrocq}),
and only derive the missing piece
regarding the derivation of the electron density perturbation.
The starting point is the electron Vlasov equation (Eq.~(8) in Ducrocq et al.~\cite{Ducrocq}),
\begin{equation}\label{eq:Vlasov_app}
\frac{\partial f_e}{\partial t}
+ \bm{v}\cdot \nabla_{\bm r} f_e
-\frac{e}{m}\Big[-\nabla \Phi + \bm{v}\times \bm{B}\Big]\cdot
\nabla_{\bm v} f_e
=0,
\end{equation}
where $\bm v$ is the electron velocity (subscript `e' dropped), $f_e$ is the electron
distribution function, $\Phi$ and $\bm B$ are the electric potential and magnetic field.

We consider a uniform crossed-field equilibrium with $\bm B=\bm B_0$ and a steady
electrostatic potential $\Phi_0(\bm r)$:
\begin{equation}\label{eq:split_fields}
\Phi(\bm r,t)=\Phi_0(\bm r)+\Phi_1(\bm r,t),\qquad
\bm E_0=-\nabla\Phi_0,\qquad
\bm B=\bm B_0,
\end{equation}
where the 1st order perturbation is denoted by subscript `1'.
The equilibrium electron distribution is assumed to be a drifting Maxwellian,
\begin{align}
f_0(\bm v)
&=n_0\left(\frac{m}{2\pi T}\right)^{3/2}
\exp\!\left[-\frac{(\bm v-\bm V_d)^2}{2V_{\rm th}^2}\right],
\\
V_{\rm th}&=\sqrt{\frac{T}{m}},\qquad \bm V_d=V_d\,\hat{\bm y}.
\end{align}
We assume an electrostatic plane-wave perturbation with complex amplitude $\tilde{\Phi}$,
\begin{equation}\label{eq:phi1_app}
\Phi_1(\bm r,t)=\tilde{\Phi}\,e^{i(\bm k\cdot\bm r-\omega t)},
\qquad
\bm E_1=-\nabla\Phi_1=-i\bm k\,\Phi_1,
\end{equation}
where $\bm{k}$ denotes the wave vector and
$\omega$ denotes the angular frequency.
We decompose the distribution function as
\begin{equation}\label{eq:fe_split_app}
f_e(\bm r,\bm v,t)=f_0(\bm v)+f_1(\bm v)\,e^{i(\bm k\cdot\bm r-\omega t)} .
\end{equation}


For a uniform magnetic field $\bm B_0=B_0\hat{\bm z}$,
thus $\bm{E}_0$ is in $-\hat{x}$, the linearized equation
for the perturbation amplitude $f_1(\bm v)$ can be written in the compact form (in the drift frame)
\begin{equation}\label{eq:lin_final}
\left[
-i\omega+i\bm k\cdot\bm v
+\Omega\frac{\partial}{\partial\theta}
\right]f_1
=-
\frac{e}{m}\,i\bm k\!\cdot\!\nabla_{\bm v} f_0\,\tilde{\Phi},
\qquad
\Omega=\frac{eB_0}{m},
\end{equation}
where $\theta$ is the gyrophase angle in the perpendicular $(x,y)$ velocity plane.
The detailed derivation from the Vlasov equation to the linearized perturbation
equation is provided in Appendix~\ref{app:linearization}.

\subsection{Retarded orbit integral solution}
\label{subsec:orbit_integral}


Next,
it is convenient to rewrite Eq.~(\ref{eq:lin_final})
as a first-order equation in the gyrophase:
\begin{equation}\label{eq:L1_theta}
\Omega\,\frac{\partial f_1}{\partial\theta}
+i\big(\bm k\!\cdot\!\bm v-\omega\big)f_1
=S(\bm v,\theta),
\qquad
S(\bm v,\theta)\equiv -\frac{e}{m}\,i\bm k\!\cdot\!\nabla_{\bm v} f_0\,\tilde{\Phi}.
\end{equation}
We solve Eq.~\eqref{eq:L1_theta} by integrating along unperturbed orbits
(method of characteristics). In a uniform magnetic field, the gyrophase evolves as
\begin{equation}\label{eq:theta_char}
\frac{d\theta}{dt}=\Omega.
\end{equation}
Introducing the look-back time $s\equiv t-t'\ge 0$ (retarded solution), the
gyrophase along the unperturbed orbit is
\begin{equation}\label{eq:theta_s}
\theta(s)=\theta-\Omega s.
\end{equation}
Define the perturbation evaluated on this orbit,
\begin{equation}\label{eq:def_g}
g(s)\equiv f_1\!\big(\bm v(s),\theta(s)\big)
=f_1\!\big(\bm v(s),\theta-\Omega s\big),
\end{equation}
where $\bm v(s)$ is the unperturbed velocity (for uniform $B_0$, $v_\perp$ and
$v_\parallel$ are constants while the perpendicular direction rotates with
$\theta(s)$). By the chain rule,
\begin{equation}\label{eq:chain_rule_g}
\frac{dg}{ds}=\frac{\partial f_1}{\partial\theta}\frac{d\theta}{ds}
=-\Omega\,\frac{\partial f_1}{\partial\theta},
\end{equation}
so that Eq.~\eqref{eq:L1_theta} becomes an ODE along the orbit,
\begin{equation}\label{eq:ode_g}
\frac{dg}{ds}-i\big(\bm k\!\cdot\!\bm v(s)-\omega\big)g(s)
=-S\!\big(\bm v(s),\theta(s)\big).
\end{equation}
Introducing the integrating factor
\begin{equation}\label{eq:int_factor_L2}
\mu(s)=\exp\!\left[-i\int_{0}^{s}\big(\bm k\!\cdot\!\bm v(\sigma)-\omega\big)\,d\sigma\right],
\end{equation}
one has
\begin{equation}\label{eq:total_derivative_L2}
\frac{d}{ds}\Big[g(s)\mu(s)\Big]
=-S\!\big(\bm v(s),\theta(s)\big)\mu(s).
\end{equation}
Imposing the retarded boundary condition $g(s\to\infty)=0$
and noting $\mu(0)=1$ yield
\begin{equation}\label{eq:L2_general}
f_1(\bm v,\theta)=g(0)
=\int_{0}^{\infty} ds\;
S\!\big(\bm v(s),\theta(s)\big)\,
\exp\!\left[-i\int_{0}^{s}\big(\bm k\!\cdot\!\bm v(\sigma)-\omega\big)\,d\sigma\right].
\end{equation}
The phase integral can be written in terms of the unperturbed orbit displacement,
\begin{equation}\label{eq:phase_displacement}
\int_{0}^{s}\bm k\!\cdot\!\bm v(\sigma)\,d\sigma
=\bm k\!\cdot\!\int_{0}^{s}\bm v(\sigma)\,d\sigma
=\bm k\!\cdot\!\big[\bm r(t)-\bm r(t-s)\big],
\end{equation}
so that
\begin{equation}\label{eq:L2_displacement_form}
f_1(\bm v,\theta)
=-\frac{e}{m}\,i\bm k\!\cdot\!\nabla_{\bm v} f_0\,\tilde{\Phi}
\int_{0}^{\infty} ds\;
\exp\!\left\{\,i\bm k\!\cdot\!\big[\bm r(t-s)-\bm r(t)\big]+i\omega s\right\}.
\end{equation}
In crossed fields, the unperturbed orbit may be decomposed as
$\bm r(t)=\bm R(t)+\bm\rho(t)$, where $\bm R$ is the guiding-center position and
$\bm\rho$ is the Larmor-radius vector. The guiding center drifts with
$\dot{\bm R}=\bm V_d+v_\parallel\hat{\bm z}$, hence
\begin{equation}\label{eq:gc_displacement}
\bm R(t-s)-\bm R(t)=-(\bm V_d+v_\parallel\hat{\bm z})\,s.
\end{equation}
Therefore,
\begin{equation}\label{eq:orbit_displacement_split}
\bm k\!\cdot\!\big[\bm r(t-s)-\bm r(t)\big]
=-(\bm k\!\cdot\!\bm V_d+k_\parallel v_\parallel)s
+\bm k_\perp\!\cdot\!\big[\bm\rho(t-s)-\bm\rho(t)\big],
\end{equation}
and Eq.~\eqref{eq:L2_displacement_form} can be cast into the standard unperturbed-orbit
integral form
\begin{equation}\label{eq:L2_final_form}
f_1(\bm v,\theta)
=-\frac{e}{m}\,i\bm k\!\cdot\!\nabla_{\bm v} f_0\,\tilde{\Phi}
\int_{0}^{\infty} ds\;
e^{i(\omega-\bm k\cdot\bm V_d-k_\parallel v_\parallel)s}\,
\exp\!\left[i\bm k_\perp\!\cdot\!\big(\bm\rho(t-s)-\bm\rho(t)\big)\right].
\end{equation}
Equation~\eqref{eq:L2_final_form} expresses the perturbation as the cumulative
wave forcing experienced by a particle traced backward along its unperturbed
orbit (retarded response).

\subsection{Finite-Larmor-radius phase and Bessel expansion}
\label{subsec:flr_bessel}

We then write the Larmor-radius vector in the $(x,y)$ plane as
\begin{equation}\label{eq:rho_def}
\bm\rho(t)=\rho_L\big(\sin\theta\,\hat{\bm x}-\cos\theta\,\hat{\bm y}\big),
\qquad
\rho_L\equiv \frac{v_\perp}{\Omega},
\end{equation}
so that, using $\theta(s)=\theta-\Omega s$,
\begin{equation}\label{eq:rho_retarded}
\bm\rho(t-s)=\rho_L\big(\sin(\theta-\Omega s)\,\hat{\bm x}-\cos(\theta-\Omega s)\,\hat{\bm y}\big).
\end{equation}
Introduce the polar angle $\alpha$ of the perpendicular wave vector,
\begin{equation}\label{eq:kperp_angle}
\bm k_\perp = k_\perp(\cos\alpha\,\hat{\bm x}+\sin\alpha\,\hat{\bm y}),
\qquad
\lambda \equiv k_\perp\rho_L=\frac{k_\perp v_\perp}{\Omega}.
\end{equation}
Then
\begin{equation}\label{eq:kdotrho}
\bm k_\perp\!\cdot\!\bm\rho(t)=\lambda\sin(\theta-\alpha),
\qquad
\bm k_\perp\!\cdot\!\bm\rho(t-s)=\lambda\sin(\theta-\Omega s-\alpha),
\end{equation}
and the orbit phase factor in Eq.~\eqref{eq:L2_final_form} becomes
\begin{align}\label{eq:phase_split}
\exp\!\left[i\bm k_\perp\!\cdot\!\big(\bm\rho(t-s)-\bm\rho(t)\big)\right]
&=\exp\!\left[i\lambda\sin(\theta-\Omega s-\alpha)\right]\,
\exp\!\left[-i\lambda\sin(\theta-\alpha)\right].
\end{align}
Using the Jacobi--Anger expansions
\begin{equation}\label{eq:JA}
e^{\,i\lambda\sin\varphi}=\sum_{n=-\infty}^{\infty}J_n(\lambda)e^{in\varphi},
\qquad
e^{-i\lambda\sin\varphi}=\sum_{m=-\infty}^{\infty}J_m(\lambda)e^{-im\varphi},
\end{equation}
we obtain the Bessel-series representation
\begin{align}\label{eq:phase_bessel_double}
\exp\!\left[i\bm k_\perp\!\cdot\!\big(\bm\rho(t-s)-\bm\rho(t)\big)\right]
&=\left[\sum_{n=-\infty}^{\infty}J_n(\lambda)e^{in(\theta-\Omega s-\alpha)}\right]
  \left[\sum_{m=-\infty}^{\infty}J_m(\lambda)e^{-im(\theta-\alpha)}\right]
\nonumber\\
&=\sum_{n=-\infty}^{\infty}\sum_{m=-\infty}^{\infty}
J_n(\lambda)J_m(\lambda)\,
e^{i(n-m)(\theta-\alpha)}\,e^{-in\Omega s}.
\end{align}

Substituting Eq.~\eqref{eq:phase_bessel_double} into Eq.~\eqref{eq:L2_final_form}
yields
\begin{align}\label{eq:f1_bessel_before_theta_avg}
f_1(\bm v,\theta)
&=-\frac{e}{m}\,i\bm k\!\cdot\!\nabla_{\bm v} f_0\,\tilde{\Phi}
\sum_{n,m=-\infty}^{\infty}J_n(\lambda)\,J_m(\lambda)\,
e^{i(n-m)(\theta-\alpha)}
\int_{0}^{\infty}ds\;
e^{\,i(\omega-\bm k\cdot\bm V_d-k_\parallel v_\parallel-n\Omega)s}.
\end{align}

\subsection{Time integral and the Landau prescription}
\label{subsec:time_landau}

Define
\begin{equation}\label{eq:Delta_n_def}
\Delta_n \equiv \omega-\bm k\!\cdot\!\bm V_d-k_\parallel v_\parallel-n\Omega .
\end{equation}
With the retarded boundary condition, we regularize the oscillatory integral by a
vanishingly small damping factor $e^{-\eta s}$ ($\eta\to 0^+$):
\begin{align}\label{eq:time_int_eval}
\int_{0}^{\infty} ds\,e^{i\Delta_n s}
&=\lim_{\eta\to 0^+}\int_{0}^{\infty} ds\,e^{i\Delta_n s-\eta s}
=\lim_{\eta\to 0^+}\frac{1}{\eta-i\Delta_n}
=\frac{i}{\Delta_n+i0^+},
\end{align}
where more detail is given in Appendix \ref{sec:prescription}.
Substituting Eq.~\eqref{eq:time_int_eval} into
Eq.~\ref{eq:f1_bessel_before_theta_avg} gives
\begin{equation}\label{eq:f1_after_time_int}
f_1(v,\theta)
=-\frac{e}{m}\,i\,\bm k\!\cdot\!\nabla_{\bm v}f_0\,\tilde{\Phi}
\sum_{n,m=-\infty}^{\infty}
J_n(\lambda)J_m(\lambda)\,e^{i(n-m)(\theta-\alpha)}
\frac{i}{\omega-\bm k\!\cdot\!\bm V_d-k_\parallel v_\parallel-n\Omega+i0^+}.
\end{equation}

\subsection{Velocity-space moments and evaluation of the density
perturbation}
\label{subsec:velocity_moments}


The perturbed electron density is obtained by taking the velocity moment of
$f_1$,
\begin{equation}\label{eq:ne1_def}
n_{e1}=\int f_1\,d^3v
=\int_{-\infty}^{\infty}dv_\parallel
\int_{0}^{\infty}dv_\perp\,v_\perp
\int_{0}^{2\pi}d\theta\; f_1(v_\perp,v_\parallel,\theta).
\end{equation}

Starting from the Bessel-expanded form of $f_1$ (Eq.~(\ref{eq:f1_after_time_int})),
\begin{equation}\label{eq:f1_bessel_start}
f_1
= \frac{e}{m}\tilde{\Phi}
\sum_{n,m=-\infty}^{\infty}
J_n(\lambda)J_m(\lambda)\,
e^{i(n-m)(\theta-\alpha)}\,
\frac{\bm k\cdot\nabla_{\bm v}f_0}{\omega-k_yV_d-k_zv_\parallel-n\Omega+i0^+}.
\end{equation}

For the (drift-frame) Maxwellian equilibrium
$f_0=f_0(v_\perp,v_\parallel)$, one has
\begin{equation}\label{eq:kdotgradf0}
\bm k\cdot\nabla_{\bm v}f_0
=\left(k_z\frac{\partial}{\partial v_\parallel}
+k_\perp\cos(\theta-\alpha)\frac{\partial}{\partial v_\perp}\right)f_0
=-\frac{f_0}{V_{\rm th}^2}\Big(k_zv_\parallel+k_\perp v_\perp\cos(\theta-\alpha)\Big).
\end{equation}
See Appendix \ref{sec:drift-maxwellian} for more detail.

Substituting \eqref{eq:kdotgradf0} into \eqref{eq:f1_bessel_start} and inserting
into \eqref{eq:ne1_def}, the $\theta$ integral involves the identities
(Appendix \ref{sec:theta})
\begin{align}
\int_{0}^{2\pi}d\theta\;e^{i(n-m)(\theta-\alpha)}
&=2\pi\,\delta_{nm},
\\
\int_{0}^{2\pi}d\theta\;\cos(\theta-\alpha)\,e^{i(n-m)(\theta-\alpha)}
&=\pi\left(\delta_{n,m+1}+\delta_{n,m-1}\right).
\end{align}
After performing the $\theta$ integral, one obtains
\begin{align}\label{eq:ne1_after_theta}
n_{e1}
&=-\frac{e}{m}\frac{\tilde{\Phi}}{V_{\rm th}^2}
\int dv_\parallel\int dv_\perp\,v_\perp\, f_0
\sum_{n=-\infty}^{\infty}
\frac{1}{\omega-k_yV_d-k_zv_\parallel-n\Omega+i0^+}
\nonumber\\
&\quad\times
\Bigg[
2\pi k_z v_\parallel J_n^2(\lambda)
+\pi k_\perp v_\perp J_n(\lambda)\Big(J_{n-1}(\lambda)+J_{n+1}(\lambda)\Big)
\Bigg].
\end{align}
Using the Bessel recursion relation
\begin{equation}
J_{n-1}(\lambda)+J_{n+1}(\lambda)=\frac{2n}{\lambda}J_n(\lambda),
\end{equation}
the bracket becomes
\[
2\pi\left(k_zv_\parallel+n\Omega\right)J_n^2(\lambda),
\]
and therefore
\begin{equation}\label{eq:ne1_theta_simplified}
n_{e1}
=-\frac{e}{m}\frac{\tilde{\Phi}}{V_{\rm th}^2}
\;2\pi\sum_{n=-\infty}^{\infty}
\int dv_\parallel\int dv_\perp\,v_\perp\, f_0\,
\frac{\left(k_zv_\parallel+n\Omega\right)J_n^2(\lambda)}
{\omega-k_yV_d-k_zv_\parallel-n\Omega+i0^+}.
\end{equation}

Introduce
\begin{equation}
\zeta_n\equiv
\frac{\omega-k_yV_d-n\Omega}{k_zV_{\rm th}\sqrt2},
\qquad
t\equiv \frac{v_\parallel}{V_{\rm th}\sqrt2},
\end{equation}
so that
\[
\omega-k_yV_d-k_zv_\parallel-n\Omega
=k_zV_{\rm th}\sqrt2(\zeta_n-t).
\]
The plasma dispersion function (Fried--Conte) is
\begin{equation}
Z(\zeta)=\frac{1}{\sqrt\pi}\int_{-\infty}^{\infty}\frac{e^{-t^2}}{t-\zeta}\,dt,
\end{equation}
the $v_\parallel$ integral in \eqref{eq:ne1_theta_simplified} yields
(Appendix \ref{sec:integral})
\begin{equation}\label{eq:vpar_integrated}
\int_{-\infty}^{\infty}dv_\parallel\,
f_0\,
\frac{k_zv_\parallel+n\Omega}{\omega-k_yV_d-k_zv_\parallel-n\Omega+i0^+}
=
-n_0\frac{e^{-v_\perp^2/(2V_{\rm th}^2)}}{2\pi V_{\rm th}^2}
\left[
1+\zeta_0 Z(\zeta_n)
\right],
\end{equation}
so that

\begin{equation}\label{eq:ne1_after_vpar}
n_{e1}
=\frac{e}{m}\frac{n_0\tilde{\Phi}}{V_{\rm th}^4}
\sum_{n=-\infty}^{\infty}
\int_0^\infty dv_\perp\,v_\perp\;
e^{-v_\perp^2/(2V_{\rm th}^2)}\,
J_n^2(\lambda)\,
\Big[\,1+\zeta_0\,Z(\zeta_n)\,\Big],
\end{equation}
where we have introduced
\begin{equation}\label{eq:zeta_defs}
\zeta_0\equiv \frac{\omega-k_yV_d}{k_zV_{\rm th}\sqrt2},
\qquad
\zeta_n\equiv \frac{\omega-k_yV_d-n\Omega}{k_zV_{\rm th}\sqrt2},
\end{equation}
so that only the plasma dispersion function $Z(\zeta_n)$ remains after the
$v_\parallel$ integration.

Next, define the magnetization parameter
\begin{equation}\label{eq:b_def}
b\equiv \frac{k_\perp^2V_{\rm th}^2}{\Omega^2},
\qquad k_\perp\equiv\sqrt{k_x^2+k_y^2},
\end{equation}
and use the standard integral (with $\lambda=k_\perp v_\perp/\Omega$)
\begin{equation}\label{eq:J2_to_I}
\int_{0}^{\infty}dv_\perp\,v_\perp\;
e^{-v_\perp^2/(2V_{\rm th}^2)}\,
J_n^2\!\left(\frac{k_\perp v_\perp}{\Omega}\right)
=
V_{\rm th}^2\,e^{-b}\,I_n(b),
\end{equation}
where $I_n$ is the modified Bessel function of the first kind,
see Appendix \ref{sec:bessel}.
Introducing the usual notation
\begin{equation}\label{eq:Gamma_n_def}
\Gamma_n(b)\equiv e^{-b}I_n(b),
\qquad \Gamma_{-n}(b)=\Gamma_n(b),
\end{equation}
Eq.~\eqref{eq:ne1_after_vpar} becomes
\begin{equation}\label{eq:ne1_Gamma_sum}
n_{e1}
=\frac{e n_0\tilde{\Phi}}{mV_{\rm th}^2}
\sum_{n=-\infty}^{\infty}
\Gamma_n(b)\Big[\,1+\zeta_0\,Z(\zeta_n)\,\Big].
\end{equation}
Using the identity $\sum_{n=-\infty}^{\infty}\Gamma_n(b)=1$, this can be written as
\begin{equation}\label{eq:ne1_split_constant}
n_{e1}
=\frac{e n_0\tilde{\Phi}}{mV_{\rm th}^2}
\left\{
1+\zeta_0\sum_{n=-\infty}^{\infty}\Gamma_n(b)\,Z(\zeta_n)
\right\}.
\end{equation}

For $n\ge 1$ it is convenient to pair the $+n$ and $-n$ harmonics. Define
\begin{equation}\label{eq:xi_pm_def}
\xi_n^\pm \equiv \frac{\omega-k_yV_d\pm n\Omega}{k_zV_{\rm th}\sqrt2},
\qquad\Rightarrow\qquad
\zeta_{-n}=\xi_n^+,\;\; \zeta_{n}=\xi_n^-,
\qquad \zeta_0=\xi_0=\frac{\xi_n^+ + \xi_n^-}{2}.
\end{equation}
Then
\begin{equation}\label{eq:Zsum_sym}
\sum_{n=-\infty}^{\infty}\Gamma_n(b)\,Z(\zeta_n)
=
\Gamma_0(b)\,Z(\xi_0)
+\sum_{n=1}^{\infty}\Gamma_n(b)\Big[Z(\xi_n^+)+Z(\xi_n^-)\Big].
\end{equation}
Substituting \eqref{eq:Zsum_sym} into \eqref{eq:ne1_split_constant} yields
\begin{equation}\label{eq:Ducrocq_eq9}
n_{e1}
=\frac{e n_0\tilde{\Phi}}{mV_{\rm th}^2}
\left\{
1+\xi_0\left[
\Gamma_0(b)\,Z(\xi_0)
+\sum_{n=1}^{\infty}\Gamma_n(b)\Big(Z(\xi_n^+)+Z(\xi_n^-)\Big)
\right]
\right\},
\end{equation}
i.e., Ducrocq et al.'s Eq.~(9), with $\Gamma_n(b)=e^{-b}I_n(b)$ and
$b=k_\perp^2V_{\rm th}^2/\Omega^2$.

\section{Connection between Ducrocq and Lafleur Forms}
\label{sec:ducrocq_lafleur_connection}

In Ducrocq \emph{et al.}~\cite{Ducrocq}, the ``full 3D'' dispersion relation is written in a
\emph{Poisson-form}: one computes $n_{e1}$ (their Eq.~(9)) and combines it with a cold-ion response in
Poisson's equation to obtain their dispersion relation
(Eq.~(11)).
In Lafleur \emph{et al.}~\cite{Lafleur}, the same physics is written in a \emph{dielectric-form}
$\hat{\epsilon}(\bm k,\omega)=0$, and the ion term is optionally taken kinetic.
Below we show that the two forms are algebraically equivalent in the present notation and coordinate
system (with $\bm B_0=B_0\hat{\bm z}$, $\bm V_d=V_d\hat{\bm y}$, $k_\parallel\equiv k_z$).

\subsection{From Poisson's equation to a dielectric function}
\label{subsec:poisson_to_dielectric}

For an electrostatic perturbation, Poisson's equation $\nabla^2\Phi_1=-\rho_1/\varepsilon_0$
gives in Fourier space
\begin{equation}
\label{eq:poisson_fourier_connection}
k^2\,\tilde{\Phi}=\frac{e}{\varepsilon_0}\left(n_{i1}-n_{e1}\right),
\qquad
k^2\equiv k_x^2+k_y^2+k_z^2.
\end{equation}
It is convenient to define species susceptibilities by
\begin{equation}
\label{eq:chi_def_connection}
\chi_s \;\equiv\; -\,\frac{q_s\,n_{s1}}{\varepsilon_0\,k^2\,\tilde{\Phi}},
\qquad (q_i=+e,\; q_e=-e),
\end{equation}
so that Eq.~\eqref{eq:poisson_fourier_connection} becomes the standard longitudinal dielectric condition
\begin{equation}
\label{eq:epsilon_def_connection}
\hat{\epsilon}(\bm k,\omega)\;\equiv\;1+\chi_i(\bm k,\omega)+\chi_e(\bm k,\omega)=0.
\end{equation}

Introduce the electron Debye length (consistent with $V_{\rm th}=\sqrt{T/m}$ used throughout)
\begin{equation}
\label{eq:lambdaDe_def_connection}
\lambda_{De}^2\equiv\frac{\varepsilon_0\,T}{n_0e^2}=\frac{\varepsilon_0\,mV_{\rm th}^2}{n_0e^2}.
\end{equation}


In the present derivation, the fully kinetic magnetized-electron density response is already obtained as
Eq.~\eqref{eq:ne1_split_constant} (or equivalently the symmetrized Ducrocq layout
Eq.~\eqref{eq:Ducrocq_eq9}). Using the susceptibility definition~\eqref{eq:chi_def_connection} with
$q_e=-e$ and the Debye length~\eqref{eq:lambdaDe_def_connection} gives immediately
\begin{equation}
\label{eq:chi_e_here}
\chi_e(\bm k,\omega)
=\frac{1}{k^2\lambda_{De}^2}
\left[
1+\zeta_0\sum_{n=-\infty}^{\infty}\Gamma_n(b)\,Z(\zeta_n)
\right],
\end{equation}
where $\zeta_0,\zeta_n$ are defined in Eq.~\eqref{eq:zeta_defs},
$b$ in Eq.~\eqref{eq:b_def}, and $\Gamma_n$ in Eq.~\eqref{eq:Gamma_n_def}.
If one prefers the paired-harmonic presentation used by Ducrocq, the sum in
Eq.~\eqref{eq:chi_e_here} can be rewritten using Eq.~\eqref{eq:Zsum_sym}.

Lafleur's Appendix uses the same dielectric structure but a different thermal-speed convention,
typically $v_{Te}\equiv\sqrt{2}\,V_{\rm th}$. With this identification,
Eq.~\eqref{eq:zeta_defs} is equivalently
\begin{equation}
\label{eq:zeta_lafleur_mapping}
\zeta_n=\frac{\omega-k_yV_d-n\Omega}{k_z v_{Te}},
\qquad
\zeta_0=\frac{\omega-k_yV_d}{k_z v_{Te}},
\end{equation}
so that the prefactor $\zeta_0$ in Eq.~\eqref{eq:chi_e_here} becomes the common Lafleur-style factor
$(\omega-k_yV_d)/(k_\parallel v_{Te})$ with $k_\parallel\equiv k_z$ in our coordinates.
Hence, the \emph{electron} part of Lafleur's dielectric function is identical to the Ducrocq electron response;
the difference is purely notational (choice of $v_{Te}$ and whether the harmonic sum is paired or left as an
integer sum).

\subsection{Ion Terms: cold-fluid (Ducrocq) versus kinetic (Lafleur)}
\label{subsec:ion_term_equiv}

Ducrocq \emph{et al.} \cite{Ducrocq} take cold, unmagnetized ions. The resulting cold-ion susceptibility is
\begin{equation}
\label{eq:chi_i_cold_connection}
\chi_i^{\rm(cold)}(\omega)=-\frac{\omega_{pi}^2}{\omega^2},
\qquad
\omega_{pi}^2\equiv\frac{n_0e^2}{\varepsilon_0 M},
\end{equation}
where $M$ denotes the ion mass.
Substituting Eqs.~\eqref{eq:chi_e_here} and \eqref{eq:chi_i_cold_connection} into
Eq.~\eqref{eq:epsilon_def_connection} yields the Ducrocq Poisson-form dispersion relation in the present
notation:
\begin{equation}
\label{eq:ducrocq_disp_here}
1-\frac{\omega_{pi}^2}{\omega^2}
+\frac{1}{k^2\lambda_{De}^2}
\left[
1+\zeta_0\sum_{n=-\infty}^{\infty}\Gamma_n(b)\,Z(\zeta_n)
\right]=0,
\end{equation}
or, after multiplying by $k^2\lambda_{De}^2$,
\begin{equation}
\label{eq:ducrocq_poisson_layout_here}
k^2\lambda_{De}^2\left(1-\frac{\omega_{pi}^2}{\omega^2}\right)
+\left[
1+\zeta_0\sum_{n=-\infty}^{\infty}\Gamma_n(b)\,Z(\zeta_n)
\right]=0,
\end{equation}
which is the same \emph{structure} as Ducrocq's full 3D relation (their Eq.~(11)), once the same
$(V_{\rm th},\zeta_n,b)$ conventions are adopted (already fixed by
Eqs.~\eqref{eq:zeta_defs}, \eqref{eq:b_def}).

In contrast, Lafleur \emph{et al.} \cite{Lafleur} use a kinetic (unmagnetized) drifting-Maxwellian ion response,
which replaces Eq.~\eqref{eq:chi_i_cold_connection} by the standard ion Landau susceptibility \cite{KrallTrivelpiece1973}
\begin{equation}
\label{eq:chi_i_kin_connection}
\chi_i^{\rm(kin)}(\bm k,\omega)
=\frac{1}{k^2\lambda_{Di}^2}\left[1+\zeta_i Z(\zeta_i)\right],
\qquad
\zeta_i\equiv\frac{\omega-\bm k\cdot\bm V_{di}}{k\,v_{Ti}},
\qquad
\lambda_{Di}^2\equiv\frac{\varepsilon_0 T_i}{n_0e^2},
\end{equation}
(with $v_{Ti}\equiv\sqrt{2T_i/M}$ and $\bm V_{di}$ the ion drift).

The plasma dispersion function satisfies the standard derivative identity
\begin{equation}
\label{eq:Zprime_identity}
Z'(\zeta)\equiv \frac{dZ}{d\zeta}
= -2\Big[\,1+\zeta Z(\zeta)\,\Big],
\end{equation}
so it is convenient to introduce
\begin{equation}
\label{eq:Z0_identity}
Z_0(\zeta)\equiv 1+\zeta Z(\zeta)=-\frac{1}{2}Z'(\zeta),
\end{equation}
and rewrite Eq.~\eqref{eq:chi_i_kin_connection} as
\begin{equation}
\label{eq:chi_i_Zprime_form}
\chi_i^{\rm(kin)}(\bm k,\omega)
=\frac{1}{k^2\lambda_{Di}^2}\,Z_0(\zeta_i)
=-\frac{1}{2k^2\lambda_{Di}^2}\,Z'(\zeta_i).
\end{equation}

Using $\omega_{pi}^2=n_0e^2/(\varepsilon_0 M)$ and $v_{Ti}^2=2T_i/M$, the ion Debye length becomes
\begin{equation}
\label{eq:lambdaDi_wpi_vti}
\lambda_{Di}^2=\frac{\varepsilon_0 T_i}{n_0e^2}
=\frac{T_i/M}{\omega_{pi}^2}
=\frac{v_{Ti}^2}{2\omega_{pi}^2},
\qquad\Rightarrow\qquad
\frac{1}{k^2\lambda_{Di}^2}=\frac{2\omega_{pi}^2}{k^2v_{Ti}^2}.
\end{equation}
Therefore, the Lafleur-style prefactor form is
\begin{equation}
\label{eq:chi_i_lafleur_style}
\chi_i^{\rm(kin)}(\bm k,\omega)
=\frac{2\omega_{pi}^2}{k^2v_{Ti}^2}\,Z_0(\zeta_i)
=-\frac{\omega_{pi}^2}{k^2v_{Ti}^2}\,Z'(\zeta_i).
\end{equation}

With this substitution, Eq.~\eqref{eq:epsilon_def_connection} becomes the Lafleur Appendix dielectric
equation. Moreover, in the cold-ion limit $kv_{Ti}\ll|\omega-\bm k\cdot\bm V_{di}|$ (i.e.\ $|\zeta_i|\gg1$),
$Z_0(\zeta_i)\simeq -1/(2\zeta_i^2)$ and Eq.~\eqref{eq:chi_i_lafleur_style} reduces to
$\chi_i^{\rm(kin)}\simeq -\omega_{pi}^2/(\omega-\bm k\cdot\bm V_{di})^2$, which recovers
Eq.~\eqref{eq:chi_i_cold_connection} when $\bm V_{di}=0$.

\section{Discussion and outlook}
\label{sec:discussion}

\subsection{Why integrate along unperturbed orbits?}
\label{subsec:why_unperturbed_orbit}

The retarded integration along unperturbed particle orbits (method of characteristics) is used
because the linearized Vlasov equation is a first-order hyperbolic equation whose natural solution
is obtained by following phase-space trajectories generated by the \emph{zeroth-order} Lorentz force.
In other words, the linear operator on the left-hand side of Eq.~\eqref{eq:lin_final} is precisely
the streaming operator associated with the equilibrium dynamics, and the perturbation $f_1$ at time
$t$ is the accumulated response to the wave forcing experienced along the past (retarded) orbit.
This construction makes causality explicit: the boundary condition $f_1(t'\to-\infty)=0$ (or,
equivalently, $g(s\to\infty)=0$ in the look-back time) leads directly to the Landau prescription
$\Delta\to\Delta+i0^+$ in the time integral, as discussed in Appendix~\ref{sec:prescription}.

Importantly, integrating along \emph{unperturbed} orbits is not an extra approximation beyond
linearization; it is the exact solution strategy for the linear problem. The unperturbed-orbit
representation does, however, rely on two practical conditions that are often implicit:
(i) the equilibrium fields are sufficiently steady and smooth that the zeroth-order trajectories are
well defined over the memory time of the response, and (ii) the forcing by the perturbation remains
small enough that orbit deviations are higher order, i.e.\ $\mathcal{O}(\Phi_1)$ corrections to the
trajectory do not feed back into $f_1$ at the same order. When these conditions fail (e.g.\ strongly
turbulent equilibria, rapidly time-varying backgrounds, or large-amplitude waves), a linear response
built on unperturbed characteristics may no longer capture the dominant transport mechanisms.

\subsection{Limitations of the Ducrocq/Lafleur assumptions in realistic devices}
\label{subsec:limitations_realistic}

The Ducrocq and Lafleur derivations share a common idealized framework: uniform crossed fields,
spatially homogeneous equilibrium parameters, electrostatic perturbations, and collisionless kinetic
electrons with FLR and cyclotron-harmonic effects retained. These assumptions are essential to obtain
a closed 3D dispersion relation, but they may be restrictive in realistic $\bm E\times\bm B$ devices
such as Hall thrusters and Penning discharges.

\paragraph{Nonuniformity and geometry.}
Real devices exhibit strong gradients in density, temperature, and electric field, as well as
magnetic-field curvature and finite channel geometry. The local (WKB) approximation implicitly
assumes the mode wavelength is short compared with equilibrium scale lengths, i.e.\ $kL\gg 1$ (or $\lambda\ll L$).
When gradients are
comparable to the wavelength, additional drift terms (e.g.\ diamagnetic effects) and profile
variations can modify both the resonance structure and the unstable spectrum.

\paragraph{Electromagnetic and boundary effects.}
The present treatment is electrostatic and assumes an unbounded or effectively periodic medium.
In experiments, conducting walls, dielectric boundaries, and sheath physics can strongly influence
wave reflection, mode structure, and energy balance. At sufficiently high frequencies or for
finite-$\beta$ regimes, electromagnetic corrections may also become non-negligible.
Here ``finite-$\beta$'' means that the plasma beta $\beta\equiv p/(B_0^2/2\mu_0)$ is not asymptotically small,
where $p$ denotes the plasma (thermal) pressure,
so magnetic-field perturbations and inductive effects can matter.

\paragraph{Collisions and sources.}
Hall thruster plasmas are partially collisional (electron--neutral, electron--ion) and include
sources and sinks (ionization, losses). Collisions can broaden resonances, alter the effective
phase relation between fields and currents, and compete with kinetic damping/driving mechanisms.
A strictly collisionless susceptibility may therefore misestimate growth rates and thresholds in
some operating regimes.

\paragraph{Ion closure and magnetization.}
Ducrocq assumes cold, unmagnetized ions, while Lafleur often uses kinetic unmagnetized ions. In
reality, ions may be warm and, depending on parameters, weakly magnetized or influenced by finite
Larmor radius and finite transit-time effects. Moreover, ion flow is typically nonuniform and may
depart from a simple drifting Maxwellian due to acceleration, ionization, and boundary losses.
These factors can change the ion resonance function and hence the coupled electron-ion instability.

\paragraph{Nonlinear saturation and anomalous transport.}
Finally, the dispersion relation is a linear theory: it predicts mode frequencies and growth rates
but not the nonlinear saturation level that ultimately determines anomalous transport. In many
devices, the turbulent state involves mode coupling, intermittency, and coherent structures, so
quantitative transport predictions require nonlinear modeling beyond the linear spectrum.

\section*{Acknowledgment}
The authors acknowledge the support from
National Natural Science Foundation of China
(Grant No.~5247120164).

\section*{Data Availability}
No new data were created or analyzed in this study.

\section*{Conflict of Interest}
On behalf of all authors, the corresponding author states that there is no
conflict of interest.

\appendix

\section{Linearization}
\label{app:linearization}

\subsection{Zeroth order: equilibrium equation}

Substituting Eqs.~\eqref{eq:split_fields} and \eqref{eq:fe_split_app} into
Eq.~\eqref{eq:Vlasov_app} and collecting the terms independent of the wave factor
$e^{i(\bm k\cdot\bm r-\omega t)}$ yields the stationary equilibrium Vlasov equation,
\begin{equation}\label{eq:equil_vlasov_app}
-\frac{e}{m}\Big[-\nabla\Phi_0+\bm v\times\bm B_0\Big]\cdot\nabla_{\bm v} f_0=0,
\end{equation}
or, equivalently,
\begin{equation}\label{eq:equil_vlasov_app2}
\Big[\bm E_0+\bm v\times\bm B_0\Big]\cdot\nabla_{\bm v}f_0=0,
\qquad \bm E_0\equiv -\nabla\Phi_0.
\end{equation}
This equation is what removes any zeroth-order contribution including
$(\bm v\times\bm B_0)\cdot\nabla_{\bm v}f_0$ from the linear response: the linear
equation is obtained after subtracting the equilibrium balance.

For the drifting Maxwellian,
\begin{equation}\label{eq:gradf0_app}
\nabla_{\bm v} f_0
=-\frac{\bm v-\bm V_d}{V_{\rm th}^2}\,f_0,
\end{equation}
so Eq.~\eqref{eq:equil_vlasov_app2} can be written as
\begin{equation}\label{eq:equil_dot_app}
\Big[\bm E_0+\bm v\times\bm B_0\Big]\cdot(\bm v-\bm V_d)=0.
\end{equation}
Expanding the dot product,
\[
\bm E_0\cdot(\bm v-\bm V_d)
+(\bm v\times\bm B_0)\cdot\bm v
-(\bm v\times\bm B_0)\cdot\bm V_d=0.
\]
Using $(\bm v\times\bm B_0)\cdot\bm v=0$ and the scalar triple product identity
$(\bm a\times\bm b)\cdot\bm c=(\bm c\times\bm b)\cdot\bm a$, we obtain
\begin{equation}\label{eq:equil_reduce_app}
\bm E_0\cdot(\bm v-\bm V_d)-(\bm V_d\times\bm B_0)\cdot\bm v=0.
\end{equation}
In a uniform crossed-field equilibrium the drift velocity is chosen as the
$\bm E_0\times\bm B_0$ drift,
\begin{equation}\label{eq:ExB_equil_app}
\bm E_0+\bm V_d\times\bm B_0=\bm 0
\qquad
(\text{equivalently } \bm V_d=\bm E_0\times\bm B_0/B_0^2),
\end{equation}
so that $(\bm V_d\times\bm B_0)\cdot\bm v=-\bm E_0\cdot\bm v$ for any $\bm v$ and
Eq.~\eqref{eq:equil_reduce_app} is satisfied identically (also $\bm E_0\cdot\bm V_d=0$).
Thus, $f_0$ is a stationary solution in uniform crossed fields, and the magnetic term
acting on $f_0$ belongs to the zeroth-order balance rather than the linear response.

\subsection{First order: linear equation}

We now retain only first-order terms in $f_1$ and $\Phi_1$.
Let $\psi\equiv \bm k\cdot\bm r-\omega t$. From Eqs.~\eqref{eq:phi1_app} and
\eqref{eq:fe_split_app},
\begin{align}
\frac{\partial f_e}{\partial t}
&=\frac{\partial}{\partial t}\!\left[f_0+f_1 e^{i\psi}\right]
=-i\omega\,f_1 e^{i\psi},
\\
\nabla_{\bm r} f_e
&=\nabla_{\bm r}\!\left[f_1 e^{i\psi}\right]
=i\bm k\,f_1 e^{i\psi},
\\
\nabla_{\bm v} f_e
&=\nabla_{\bm v} f_0+\left(\nabla_{\bm v} f_1\right)e^{i\psi},
\\
-\nabla\Phi
&=-\nabla(\Phi_0+\Phi_1)
=\bm E_0 - i\bm k\,\tilde{\Phi}\,e^{i\psi}.
\end{align}
Substituting into Eq.~\eqref{eq:Vlasov_app} and discarding the second-order term
$\mathcal{O}(\tilde{\Phi}f_1)$ (proportional to $e^{2i\psi}$), we collect all
remaining first-order terms (proportional to $e^{i\psi}$) and divide by $e^{i\psi}$.
This yields
\begin{equation}\label{eq:lin_pre_app}
\left[
-i\omega+i\bm k\cdot\bm v
-\frac{e}{m}\Big(\bm E_0+\bm v\times\bm B_0\Big)\cdot\nabla_{\bm v}
\right]f_1
=
-\frac{e}{m}\,i\bm k\cdot\nabla_{\bm v} f_0\,\tilde{\Phi}.
\end{equation}

Finally, for $\bm B_0=B_0\hat{\bm z}$, one uses the standard identity
\begin{equation}\label{eq:gyro_identity}
-\frac{e}{m}\,(\bm v\times\bm B_0)\cdot\nabla_{\bm v}
=\Omega\,\frac{\partial}{\partial\theta},
\qquad
\Omega=\frac{eB_0}{m},
\end{equation}
whose derivation is given in Appendix~\ref{app:gyro_identity}.
Using Eq.~\eqref{eq:gyro_identity}, Eq.~\eqref{eq:lin_pre_app} becomes
\begin{equation}\label{eq:linear_eq_f1_app}
\left[
-i\omega+i\bm k\cdot\bm v
+\Omega\frac{\partial}{\partial\theta}
-\frac{e}{m}\bm E_0\cdot\nabla_{\bm v}
\right]f_1
=
-\frac{e}{m}\,i\bm k\cdot\frac{\partial f_0}{\partial\bm v}\,\tilde{\Phi}.
\end{equation}
In the drift frame (equivalently, after absorbing the uniform $\bm E_0\times\bm B_0$
drift into the Doppler shift $\omega\to\omega-\bm k\cdot\bm V_d$), the term
$-(e/m)\bm E_0\cdot\nabla_{\bm v}$ can be eliminated, yielding Eq.~\eqref{eq:lin_final}.

In more detail,
the uniform crossed fields $\bm E_0\perp \bm B_0$ produce a constant
$\bm E_0\times\bm B_0$ drift of the guiding center. Changing to the drift frame
\begin{equation}
\bm r'=\bm r-\bm V_d t,\qquad
\bm v'=\bm v-\bm V_d,
\end{equation}
one finds
\begin{equation}
\bm E_0+\bm v\times\bm B_0
=\bm E_0+(\bm v'+\bm V_d)\times\bm B_0
=\underbrace{\big(\bm E_0+\bm V_d\times\bm B_0\big)}_{=\bm 0}
+\bm v'\times\bm B_0
=\bm v'\times\bm B_0.
\end{equation}
Since $\nabla_{\bm v}=\nabla_{\bm v'}$ under a constant velocity shift, the
operator $-(e/m)\bm E_0\cdot\nabla_{\bm v}$ can thus be absorbed into the
magnetic part of the Lorentz operator in the drift frame.
Equivalently, the plane-wave phase transforms as
\[
e^{i(\bm k\cdot\bm r-\omega t)}
=e^{i[\bm k\cdot(\bm r'+\bm V_d t)-\omega t]}
=e^{i(\bm k\cdot\bm r'-(\omega-\bm k\cdot\bm V_d)t)},
\]
which shows that the drift-frame frequency is Doppler shifted according to
\begin{equation}
\omega\ \to\ \omega-\bm k\cdot\bm V_d.
\end{equation}

\subsection{Gyro-angle identity}\label{app:gyro_identity}

For a uniform magnetic field $\bm B_0=B_0\hat{\bm z}$, introduce cylindrical
coordinates in velocity space,
\begin{equation}
v_x=v_\perp\cos\theta,\qquad
v_y=v_\perp\sin\theta,\qquad
v_z=v_\parallel,
\end{equation}
so that $\theta$ is the gyrophase angle in the perpendicular $(x,y)$ plane.
The Lorentz operator associated with the magnetic force is
\begin{equation}
\mathcal{L}_B \equiv -\frac{e}{m}(\bm v\times\bm B_0)\cdot\nabla_{\bm v}.
\end{equation}
Since $\bm v\times\bm B_0=(v_yB_0,\,-v_xB_0,\,0)$, one has
\begin{equation}
(\bm v\times\bm B_0)\cdot\nabla_{\bm v}
=B_0\left(v_y\frac{\partial}{\partial v_x}-v_x\frac{\partial}{\partial v_y}\right).
\end{equation}
On the other hand, the gyrophase derivative at fixed $(v_\perp,v_\parallel)$ is
\begin{equation}
\frac{\partial}{\partial\theta}
=\frac{\partial v_x}{\partial\theta}\frac{\partial}{\partial v_x}
+\frac{\partial v_y}{\partial\theta}\frac{\partial}{\partial v_y}
=(-v_\perp\sin\theta)\frac{\partial}{\partial v_x}
+(v_\perp\cos\theta)\frac{\partial}{\partial v_y}
=-v_y\frac{\partial}{\partial v_x}+v_x\frac{\partial}{\partial v_y}.
\end{equation}
Combining the last two relations gives
\begin{equation}
(\bm v\times\bm B_0)\cdot\nabla_{\bm v}
=-B_0\,\frac{\partial}{\partial\theta},
\end{equation}
and therefore
\begin{equation}
-\frac{e}{m}(\bm v\times\bm B_0)\cdot\nabla_{\bm v}
=\frac{eB_0}{m}\,\frac{\partial}{\partial\theta}
\equiv \Omega\,\frac{\partial}{\partial\theta},
\qquad \Omega=\frac{eB_0}{m},
\end{equation}
which is Eq.~\eqref{eq:gyro_identity}.

Physically, Eq.~\eqref{eq:gyro_identity} expresses that the magnetic part of
the Vlasov operator generates a rotation in the perpendicular velocity space,
i.e., the electron gyromotion at the cyclotron frequency.

\section{Causality Prescription}\label{sec:prescription}

For real
$\Delta_n$, the integrand $e^{i\Delta_n s}$ is purely oscillatory and does not
decay as $s\to\infty$, so the improper integral
$\int_{0}^{\infty} ds\,e^{i\Delta_n s}$ is not convergent in the ordinary sense.
Physically, the solution must satisfy causality (the retarded condition): the perturbation at time $t$ can only depend on the past forcing along the unperturbed orbit ($t'\le t$, i.e.\ $s=t-t'\ge 0$), which is enforced by the Landau prescription $\Delta\to\Delta+i0^+$ in the time integral.
This retarded boundary condition is enforced by
introducing an infinitesimal damping factor $e^{-\eta s}$ with $\eta>0$,
evaluating the convergent integral, and finally taking the limit
$\eta\to0^+$:
\begin{align}\label{eq:retarded_detail}
\int_{0}^{\infty} ds\,e^{i\Delta_n s}
&\equiv
\lim_{\eta\to0^+}\int_{0}^{\infty} ds\,e^{i\Delta_n s-\eta s}
=
\lim_{\eta\to0^+}\int_{0}^{\infty} ds\,e^{-(\eta-i\Delta_n)s}
\nonumber\\
&=
\lim_{\eta\to0^+}\left[\frac{-1}{\eta-i\Delta_n}\,e^{-(\eta-i\Delta_n)s}\right]_{0}^{\infty}
=
\lim_{\eta\to0^+}\frac{1}{\eta-i\Delta_n}
=
\frac{i}{\Delta_n+i0^+}.
\end{align}
Here
\begin{equation}\label{eq:Delta_n_text}
\Delta_n \equiv \omega-\bm k\!\cdot\!\bm V_d-k_\parallel v_\parallel-n\Omega
\end{equation}
is the Doppler-shifted cyclotron-resonance detuning. The notation $+i0^+$
(Landau prescription) indicates that the pole at $\Delta_n=0$ is approached from
the upper half of the complex plane, which is the analytic continuation
consistent with causality.

\section{Maxwellian Gradients}\label{sec:drift-maxwellian}

For a drift-frame Maxwellian equilibrium $f_0=f_0(v_\perp,v_\parallel)$ that is
gyrotropic (i.e., independent of the gyrophase $\theta$), the velocity-space
gradient can be written in cylindrical coordinates $(v_\perp,\theta,v_\parallel)$.
In these coordinates,
\begin{equation}
\nabla_{\bm v}
=\hat{\bm e}_{v_\perp}\frac{\partial}{\partial v_\perp}
+\hat{\bm e}_{\theta}\frac{1}{v_\perp}\frac{\partial}{\partial \theta}
+\hat{\bm e}_{\parallel}\frac{\partial}{\partial v_\parallel},
\end{equation}
and since $f_0$ is gyrotropic, $\partial f_0/\partial\theta=0$.  Decompose the
wavevector into parallel and perpendicular parts,
\begin{equation}
\bm k=k_\parallel\hat{\bm z}+\bm k_\perp,
\qquad
\bm k_\perp=k_\perp(\cos\alpha\,\hat{\bm x}+\sin\alpha\,\hat{\bm y}),
\end{equation}
and note that the perpendicular unit vector is
\begin{equation}
\hat{\bm e}_{v_\perp}=\cos\theta\,\hat{\bm x}+\sin\theta\,\hat{\bm y}.
\end{equation}
Therefore,
\begin{equation}
\bm k\cdot\nabla_{\bm v} f_0
=k_\parallel\frac{\partial f_0}{\partial v_\parallel}
+\bm k_\perp\cdot\hat{\bm e}_{v_\perp}\frac{\partial f_0}{\partial v_\perp}
=
\left(
k_\parallel\frac{\partial}{\partial v_\parallel}
+k_\perp\cos(\theta-\alpha)\frac{\partial}{\partial v_\perp}
\right)f_0,
\end{equation}
where we used
$\bm k_\perp\cdot\hat{\bm e}_{v_\perp}
=k_\perp(\cos\alpha\cos\theta+\sin\alpha\sin\theta)=k_\perp\cos(\theta-\alpha)$.

For the drifting Maxwellian in the drift frame,
\begin{equation}
f_0(v_\perp,v_\parallel)
=n_0\left(\frac{m}{2\pi T}\right)^{3/2}
\exp\!\left[-\frac{v_\perp^2+v_\parallel^2}{2V_{\rm th}^2}\right],
\qquad
V_{\rm th}=\sqrt{\frac{T}{m}},
\end{equation}
one has the simple derivatives
\begin{equation}
\frac{\partial f_0}{\partial v_\parallel}
=-\frac{v_\parallel}{V_{\rm th}^2}f_0,
\qquad
\frac{\partial f_0}{\partial v_\perp}
=-\frac{v_\perp}{V_{\rm th}^2}f_0.
\end{equation}
Substituting into the previous expression yields
\begin{equation}
\bm k\cdot\nabla_{\bm v}f_0
=-\frac{f_0}{V_{\rm th}^2}\left(k_\parallel v_\parallel
+k_\perp v_\perp\cos(\theta-\alpha)\right),
\end{equation}
which is Eq.~(\ref{eq:kdotgradf0}).

\section{Gyrophase Integrals} \label{sec:theta}


In the evaluation of the velocity integrals, one repeatedly encounters the
gyrophase factor $e^{i(n-m)(\theta-\alpha)}$, where $\theta$ is the particle
gyrophase and $\alpha$ is the polar angle of the perpendicular wave vector
$\bm k_\perp=k_\perp(\cos\alpha\,\hat{\bm x}+\sin\alpha\,\hat{\bm y})$. The
following two identities are used:
\begin{align}
\int_{0}^{2\pi}\! d\theta\; e^{i(n-m)(\theta-\alpha)}
&=2\pi\,\delta_{nm}, \label{eq:theta_id_1}\\
\int_{0}^{2\pi}\! d\theta\; \cos(\theta-\alpha)\,e^{i(n-m)(\theta-\alpha)}
&=\pi\left(\delta_{n,m+1}+\delta_{n,m-1}\right). \label{eq:theta_id_2}
\end{align}
They follow directly from Fourier orthogonality.

\paragraph{Proof of Eq.~\eqref{eq:theta_id_1}.}
Let $\varphi\equiv\theta-\alpha$. Since $\alpha$ is a constant independent of
$\theta$, $d\varphi=d\theta$ and shifting the integration interval by a constant
does not change the value:
\begin{equation}
\int_{0}^{2\pi}\! d\theta\; e^{i(n-m)(\theta-\alpha)}
=\int_{-\alpha}^{2\pi-\alpha}\! d\varphi\; e^{i(n-m)\varphi}
=\int_{0}^{2\pi}\! d\varphi\; e^{i(n-m)\varphi}.
\end{equation}
For integers $n,m$, this integral is
\begin{equation}
\int_{0}^{2\pi}\! d\varphi\; e^{i(n-m)\varphi}
=
\begin{cases}
2\pi, & n=m,\\[2pt]
\dfrac{e^{i(n-m)2\pi}-1}{i(n-m)}=0, & n\neq m,
\end{cases}
\end{equation}
because $e^{i(n-m)2\pi}=1$. This is precisely $2\pi\delta_{nm}$.

\paragraph{Proof of Eq.~\eqref{eq:theta_id_2}.}
Using $\cos\varphi=\tfrac12\left(e^{i\varphi}+e^{-i\varphi}\right)$ and the same
change of variable $\varphi=\theta-\alpha$, we obtain
\begin{align}
\int_{0}^{2\pi}\! d\theta\; \cos(\theta-\alpha)\,e^{i(n-m)(\theta-\alpha)}
&=\int_{0}^{2\pi}\! d\varphi\; \cos\varphi\,e^{i(n-m)\varphi}\nonumber\\
&=\frac12\int_{0}^{2\pi}\! d\varphi\;
\left(e^{i\varphi}+e^{-i\varphi}\right)e^{i(n-m)\varphi}\nonumber\\
&=\frac12\int_{0}^{2\pi}\! d\varphi\;
\left(e^{i(n-m+1)\varphi}+e^{i(n-m-1)\varphi}\right).
\end{align}
Applying Eq.~\eqref{eq:theta_id_1} to each exponential term gives
\begin{equation}
\frac12\left[2\pi\,\delta_{n,m-1}+2\pi\,\delta_{n,m+1}\right]
=
\pi\left(\delta_{n,m+1}+\delta_{n,m-1}\right),
\end{equation}
which proves Eq.~\eqref{eq:theta_id_2}.

Equation~\eqref{eq:theta_id_1} expresses the orthogonality of Fourier harmonics
in the gyrophase angle: only the ``diagonal'' term $n=m$ survives the
$\theta$-average. Equation~\eqref{eq:theta_id_2} shows that multiplying by
$\cos(\theta-\alpha)$ couples neighboring harmonics, shifting the index by
$\pm1$; this is the origin of the $n\leftrightarrow n\pm1$ coupling that later
reduces to Bessel-function recursion relations.

\section{Parallel Integral}\label{sec:integral}

We derive the $v_\parallel$ integral used in Eq.~\eqref{eq:vpar_integrated}. The quantity
to be evaluated is
\begin{equation}\label{eq:app_vpar_target}
\mathcal{I}_n(v_\perp)\;\equiv\;
\int_{-\infty}^{\infty} dv_\parallel\;
f_0(v_\perp,v_\parallel)\,
\frac{k_z v_\parallel+n\Omega}{\omega-k_yV_d-k_z v_\parallel-n\Omega+i0^+}.
\end{equation}

For a drift-frame Maxwellian,
\begin{equation}\label{eq:app_f0_factor}
f_0(v_\perp,v_\parallel)
=n_0\left(\frac{m}{2\pi T}\right)^{3/2}
\exp\!\left[-\frac{v_\perp^2+v_\parallel^2}{2V_{\rm th}^2}\right]
=
\frac{n_0}{(2\pi)^{3/2}V_{\rm th}^3}\,
e^{-v_\perp^2/(2V_{\rm th}^2)}\,e^{-v_\parallel^2/(2V_{\rm th}^2)} .
\end{equation}
Hence the factor $e^{-v_\perp^2/(2V_{\rm th}^2)}$ can be taken outside the
$v_\parallel$ integral.

Introduce the dimensionless variables
\begin{equation}\label{eq:app_zeta_defs}
t\equiv \frac{v_\parallel}{\sqrt{2}\,V_{\rm th}},
\qquad
\zeta_n\equiv \frac{\omega-k_yV_d-n\Omega}{\sqrt{2}\,k_zV_{\rm th}},
\qquad
\zeta_0\equiv \frac{\omega-k_yV_d}{\sqrt{2}\,k_zV_{\rm th}},
\end{equation}
so that
\begin{equation}\label{eq:app_denom}
\omega-k_yV_d-k_z v_\parallel-n\Omega+i0^+
=\sqrt{2}\,k_zV_{\rm th}\,(\zeta_n-t)+i0^+.
\end{equation}
The Fried--Conte plasma dispersion function is defined as
\begin{equation}\label{eq:app_Z_def}
Z(\zeta)=\frac{1}{\sqrt{\pi}}
\int_{-\infty}^{\infty}\frac{e^{-t^2}}{t-\zeta}\,dt,
\end{equation}
with the Landau prescription implied by $+i0^+$,
i.e.\ the limit
$\lim_{\epsilon\to0^+}(t-\zeta+i\epsilon)^{-1}$, which fixes how the pole is
bypassed and yields the standard decomposition into a principal-value part and a
$\delta$-function (resonant) contribution.

Using $A\equiv \omega-k_yV_d-k_z v_\parallel-n\Omega$, we note that
\begin{equation}\label{eq:app_numer_decomp}
k_z v_\parallel+n\Omega
=(\omega-k_yV_d)-A,
\end{equation}
which immediately gives the identity
\begin{equation}\label{eq:app_key_identity}
\frac{k_z v_\parallel+n\Omega}{\omega-k_yV_d-k_z v_\parallel-n\Omega+i0^+}
=
-1+\frac{\omega-k_yV_d}{\omega-k_yV_d-k_z v_\parallel-n\Omega+i0^+}.
\end{equation}
This splits $\mathcal{I}_n$ into an elementary Maxwellian moment and a dispersive
(Landau) contribution.

Substituting Eq.~\eqref{eq:app_key_identity} into Eq.~\eqref{eq:app_vpar_target} and
using Eq.~\eqref{eq:app_f0_factor}, we obtain
\begin{equation}\label{eq:app_split}
\mathcal{I}_n(v_\perp)
=
\frac{n_0}{(2\pi)^{3/2}V_{\rm th}^3}\,e^{-v_\perp^2/(2V_{\rm th}^2)}
\left[
-\int_{-\infty}^{\infty} dv_\parallel\,e^{-v_\parallel^2/(2V_{\rm th}^2)}
+(\omega-k_yV_d)\,\mathcal{J}_n
\right],
\end{equation}
where
\begin{equation}\label{eq:app_Jn_def}
\mathcal{J}_n \equiv
\int_{-\infty}^{\infty} dv_\parallel\;
\frac{e^{-v_\parallel^2/(2V_{\rm th}^2)}}
{\omega-k_yV_d-k_z v_\parallel-n\Omega+i0^+}.
\end{equation}

The first integral is elementary,
\begin{equation}\label{eq:app_gauss}
\int_{-\infty}^{\infty} dv_\parallel\,e^{-v_\parallel^2/(2V_{\rm th}^2)}
=\sqrt{2\pi}\,V_{\rm th}.
\end{equation}

For $\mathcal{J}_n$, use the change of variables $v_\parallel=\sqrt2 V_{\rm th}t$ and
Eqs.~\eqref{eq:app_zeta_defs}--\eqref{eq:app_denom}:
\begin{align}
\mathcal{J}_n
&=\sqrt2 V_{\rm th}\int_{-\infty}^{\infty}dt\;
\frac{e^{-t^2}}{\sqrt2\,k_zV_{\rm th}(\zeta_n-t)+i0^+}
=\frac{1}{k_z}\int_{-\infty}^{\infty}dt\;\frac{e^{-t^2}}{\zeta_n-t+i0^+}
\nonumber\\
&=-\frac{1}{k_z}\int_{-\infty}^{\infty}dt\;\frac{e^{-t^2}}{t-\zeta_n-i0^+}
=-\frac{\sqrt{\pi}}{k_z}\,Z(\zeta_n).
\label{eq:app_Jn_eval}
\end{align}

Substituting Eqs.~\eqref{eq:app_gauss} and \eqref{eq:app_Jn_eval} into
Eq.~\eqref{eq:app_split}, and using $\omega-k_yV_d=\sqrt2\,k_zV_{\rm th}\,\zeta_0$,
we finally obtain
\begin{equation}\label{eq:app_vpar_result}
\mathcal{I}_n(v_\perp)
=
-\frac{n_0}{2\pi V_{\rm th}^2}\,
e^{-v_\perp^2/(2V_{\rm th}^2)}
\Big[1+\zeta_0 Z(\zeta_n)\Big].
\end{equation}
Equation~\eqref{eq:app_vpar_result} is the result quoted in
Eq.~\eqref{eq:vpar_integrated}.

\section{Perpendicular Integral} \label{sec:bessel}

We justify the standard integral used in Eq.~(\ref{eq:J2_to_I}),
\begin{equation}\label{eq:app_J2_standard}
\int_{0}^{\infty}dv_\perp\,v_\perp\;
e^{-v_\perp^2/(2V_{\rm th}^2)}\,
J_n^2\!\left(\frac{k_\perp v_\perp}{\Omega}\right)
=
V_{\rm th}^2\,e^{-b}\,I_n(b),
\qquad
b\equiv \frac{k_\perp^2V_{\rm th}^2}{\Omega^2},
\end{equation}
where $I_n$ is the modified Bessel function of the first kind.

Introduce the dimensionless variable $x\equiv v_\perp/V_{\rm th}$ and
$a\equiv k_\perp V_{\rm th}/\Omega$, so that $b=a^2$ and
$dv_\perp\,v_\perp=V_{\rm th}^2\,x\,dx$. Then Eq.~\eqref{eq:app_J2_standard}
becomes
\begin{equation}\label{eq:app_dimless}
\int_{0}^{\infty}dx\;x\,e^{-x^2/2}\,J_n^2(ax)
=
e^{-a^2}\,I_n(a^2).
\end{equation}

A standard identity (often referred to as the Weber--Sonine formula) states that,
for $\Re(p)>0$,
\begin{equation}\label{eq:app_weber_sonine}
\int_{0}^{\infty}dx\;x\,e^{-p x^2}\,J_n(\alpha x)\,J_n(\beta x)
=
\frac{1}{2p}\exp\!\left[-\frac{\alpha^2+\beta^2}{4p}\right]\,
I_n\!\left(\frac{\alpha\beta}{2p}\right).
\end{equation}
Setting $\alpha=\beta=a$, and $p=\tfrac12$ gives
\begin{equation}
\int_{0}^{\infty}dx\;x\,e^{-x^2/2}\,J_n^2(ax)
=
\exp(-a^2)\,I_n(a^2),
\end{equation}
which is precisely Eq.~\eqref{eq:app_dimless}.

Returning to $a=k_\perp V_{\rm th}/\Omega$ and $b=a^2$ yields
Eq.~\eqref{eq:app_J2_standard}, i.e.,
\[
\int_{0}^{\infty}dv_\perp\,v_\perp\;
e^{-v_\perp^2/(2V_{\rm th}^2)}\,
J_n^2\!\left(\frac{k_\perp v_\perp}{\Omega}\right)
=
V_{\rm th}^2\,e^{-b}\,I_n(b),
\qquad b=\frac{k_\perp^2V_{\rm th}^2}{\Omega^2}.
\]

\section*{References}
\printbibliography[heading=none]

\end{document}